\documentclass[%
 aip,
 amsmath,amssymb,
 reprint,%
]{revtex4-1}

\usepackage{graphicx}
\usepackage{dcolumn}
\usepackage{bm}
\usepackage{textgreek}
\usepackage[utf8]{inputenc}
\usepackage[T1]{fontenc}
\usepackage{mathptmx}
\usepackage{etoolbox}
\usepackage{csquotes}
\usepackage[usenames,dvipsnames]{xcolor} 
\usepackage{amsmath,MnSymbol}
\usepackage{subcaption}
\usepackage{lipsum}
\usepackage{empheq}

\makeatletter
\def\@email#1#2{%
 \endgroup
 \patchcmd{\titleblock@produce}
 {\frontmatter@RRAPformat}
 {\frontmatter@RRAPformat{\produce@RRAP{*#1\href{mailto:#2}{#2}}}\frontmatter@RRAPformat}
 {}{}
}%
\makeatother
\renewcommand*{\d}{\mathop{}\!\mathrm{d}}

\newcommand{\hide}[1]{{\hphantom{}}}

\begin{document}

\preprint{AIP/123-QED}

\title{Comparative Analysis of Fluctuations in Viscoelastic Stress: A Comparison of the Temporary Network and Dumbbell models}

\author{Arturo Winters}
\author{Hans Christian Öttinger}%

\affiliation{Department of Materials, ETH Zurich, Vladimir-Prelog-Weg 5, Zurich 8093, Switzerland}%
\email{jan.vermant@mat.ethz.ch}
\author{Jan Vermant}
\affiliation{%
Department of Materials, ETH Zurich, Vladimir-Prelog-Weg 5, Zurich 8093, Switzerland}%

\date{\today}

\begin{abstract}
\textbf{Abstract.} 
Traditionally, stress fluctuations in flowing and deformed materials are overlooked, with an obvious focus on average stresses in a continuum mechanical approximation. However, these fluctuations, often dismissed as \enquote{noise}, hold the potential to provide direct insights into the material structure and its structure-stress coupling, uncovering detailed aspects of fluid transport and relaxation behaviors. Despite advancements in experimental techniques allowing for the visualization of these fluctuations, their significance remains largely untapped, as modeling efforts continue to target Newtonian fluids within the confines of Gaussian noise assumptions.  In the present work  a comparative analysis of stress fluctuations in  two distinct microstructural models is carried out: the temporary network model and the dumbbell model. Despite both models conforming to the Upper Convected Maxwell Model at a macroscopic level, the  temporary network model predicts non-Gaussian fluctuations. We find that stress fluctuations within the temporary network model exhibit more pronounced abruptness at local scale, with only an enlargement of the control volume leading to a gradual Gaussian-like noise, diminishing the differences between the two models. These findings underscore the heightened sensitivity of  \emph{fluctuating rheology} to microstructural details and the microstructure-flow coupling, beyond what is captured by macroscopically averaged stresses.
\end{abstract}

\maketitle

\section{\label{sec. 1}Introduction\protect\\}
\begin{figure}%
\centering
\includegraphics[width=0.46\textwidth]{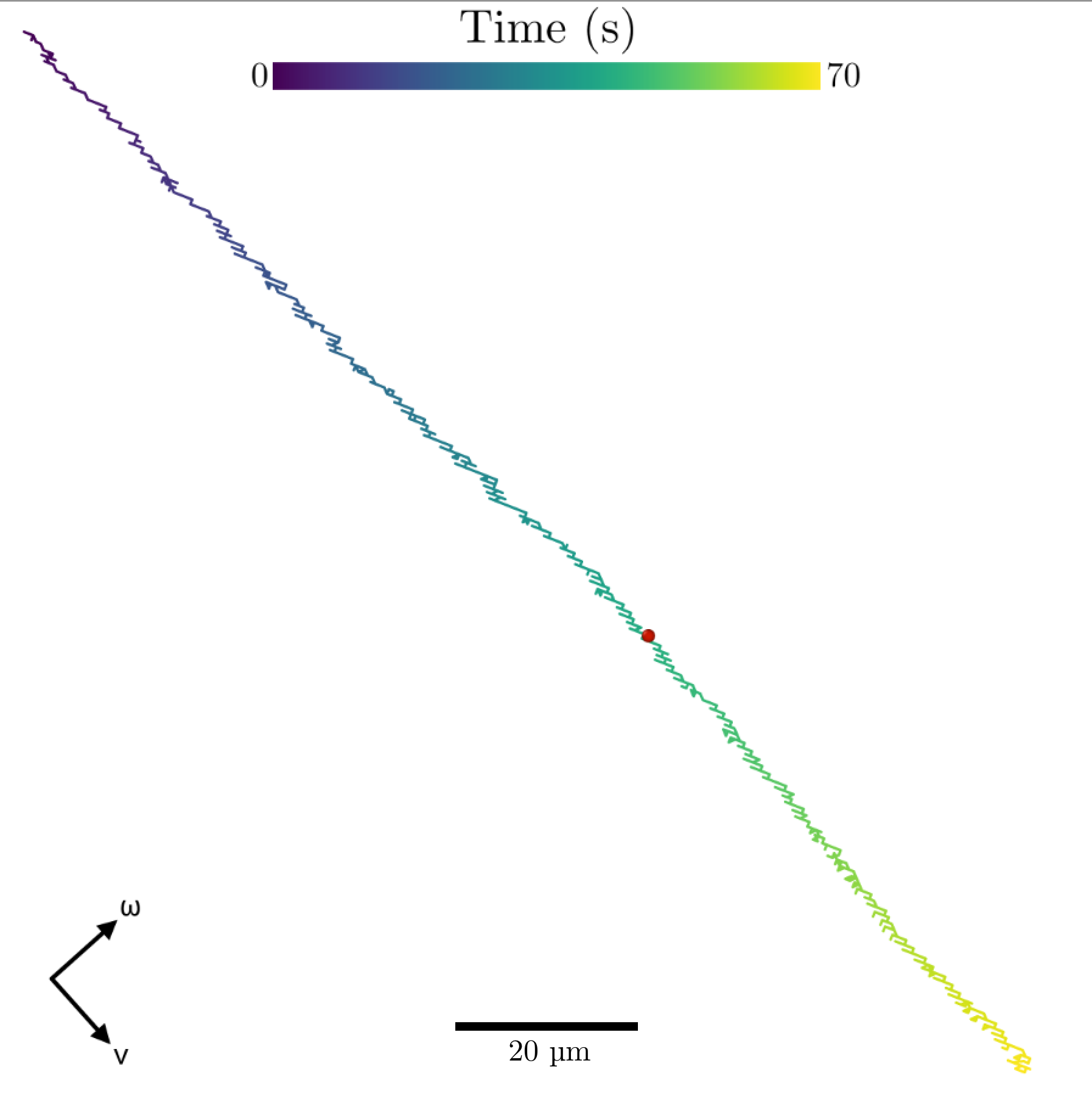}
 \caption{Tracked 1-\textmu m beads near the stagnation plane ($v,\,\omega$ for the velocity and vorticity directions, respectively) in a shear flow at rate $\dot{\gamma}=1\text{s}^{-1}$ wormlike micellar solution observed by high speed rheo-confocal microscopy.\cite{Confocal,Trackpy} The (near Maxwellian) solution at 25° characterized in \citet{Avino2009} has a single relaxation time $\lambda=1.5$ s and a modulus of $G=33.4$ Pa. The tracks show a convective motion along the flow direction with superimposed fluctuations.}
 \label{fig: Rheoconf}%
\end{figure}

In 1827, Brown\cite{Brown1827} discovered Brownian motion, a phenomenon that lies at the heart of understanding transport phenomena in soft matter and motivated Einstein's doctoral thesis to look at fluctuations.\cite{Einstein}  Einstein proposed a way to determine the size of molecules in a liquid which led to a method for calculating Avogadro's number, laying the groundwork for modern physical chemistry and statistical physics. The present paper presents  insights into thermal fluctuations, but now in flowing complex materials, using the viscoelastic Maxwell model as a fundamental rheological example. In the field of material science, thermal fluctuations at equilibrium are crucial for interpreting scattering experiments \cite{Sengers, Larsen2008, Transport2018} or and microrheological experiments,\cite{Einstein,Mason1995_1, Furst} providing a deeper understanding of material behavior at the microscopic level. Furthermore, many models of complex materials such as bead-spring models\cite{Stochastic} are based on fluctuations at a microscopic scale. To date, the role of Brownian motion in transport processes at nanoscale remains an active field of research.\cite{Marbach_1,Marbach_2}\\ \indent
The continuum mechanical modeling of fluctuations at equilibrium, rely on the coarse grained description of Landau and Lipshitz,\cite{Landau87} which treats dissipative fluxes $\pmb{J}$, e.g. the heat flux or the extra stress tensor $\pmb{\tau}$, as stochastic variables. The latter quantity is additively split into an average value and a fluctuating part $\delta\pmb{J}$. The fluctuations at equilibrium fluctuation are assumed to be random and uncorrelated at different points in space and moments in time. The second moment of  $\delta\pmb{J}$ is hence delta-correlated in space and time 
\begin{equation} \label{eq: Fluct_hydrodynamics}
\langle\delta\pmb{J}_{\alpha}\delta\pmb{J}_{\beta}\rangle=2k_{\text{B}}TK_{\alpha\beta}\delta(\pmb{r}-\pmb{r'})\delta(t-t'),
\end{equation}
where $k_{\text{B}}$ is the Boltzmann constant, $T$ the temperature, and $K_{\alpha\beta}$ are components of the so called Onsager dissipation matrix, which are reflecting material properties such as the thermal conductivity or the viscosity. The delta-correlation in time  reflects the Gaussian noise, when the system is at equilibrium. When we assume local but not global equilibrium, we can develop fluctuating hydrodynamics, which successfully models fluctuations in flows or with thermal gradients, in Newtonian fluids.\cite{Sengers, Sengers_paper} So far previous work has mainly focused on Newtonian materials, in the present work we turn to complex fluids, where the rheological properties of the fluids are more complicated, and the fluids are described by either a structural variable or \enquote{memory} effects. Extending this is motivated by recent experimental developments as, for instance, by high speed confocal counterrotating rheo-microscopy\cite{Confocal} where particle tracking  methods can be applied to flowing systems. An example is shown in fig. \ref{fig: Rheoconf}, where the tracks of micrometer sized particles fluctuate during flow of a single relaxation time (near Maxwellian) wormlike micellar solution characterized in Snijkers et al.\cite{Avino2009} The tracks show the convective motion of the particles with superimposed noise in the displacement, resulting from the combined effect of stress fluctuations on the particle.

The present works seeks to understand whether the fluctuations of the stress in complex fluids can be used to infer more detailed information on microscopic dynamics. 
The few works dealing with the topic,\cite{Huetter2020,Grmela_Oe2} as we explore further in sec. \ref{subsec: Fluctuation dissipation}, confined themselves to model fluctuations of the Gaussian type as in eq. (\ref{eq: Fluct_hydrodynamics}). Our initial objective is to advance the concept of fluctuating rheology. We begin by demonstrating that, even in the context of the simplest viscoelastic model, the Upper Convected Maxwell Model (UCM), Gaussian approximations are not universally applicable. 
Our second objective specifically addresses the nuances of the UCM, adhering to the principle that \enquote{the main value of this model is as a guide for qualitative thinking}.\cite{Ferry1980}
Remarkably, both the Temporary Network Model (TNM) and the Hydrodynamic Dumbbell Model (HDM), which serve as basic constitutive model systems for two distinct groups of microscopic models, reproduce the Maxwellian behavior.\cite{Larson,Stochastic} Till now, literature emphasized  the equality of the two models when describing the stress evolution,\cite{Ferry1980, Morrison2001,Transport2018,Larson} which is intriguing given their fundamental differences on a microscopic scale. In this paper we demonstrate how  differences do come out when considering the stress fluctuations and suggest methods to measure and use them.\\ \indent

The paper is organized as follows: First (sec. \ref{subsec: Models}), we describe the specifics of the two approaches and how they are cross-grained to yield the UCM.
Thereafter, we introduce a thermodynamic framework to express fluctuations (sec. \ref{subsec: Fluctuation dissipation}) and focus on the Fluctuation-Dissipation Theorem (FDT) and its role in connecting the microscopic world with the macroscopic one.\cite{Oettinger_1} From sec. \ref{sec: Potentials} onward we present novel results. We start with the TNM and HDM dissipation potentials (sec. \ref{sec: Potentials}). The differences of the two models with respect to the probability density $p$ and to $\pmb{\tau}$ follow in sec. \ref{sec: PDF} and \ref{sec: Fluctuations comparison}. In sec. \ref{sec: Discussion} we present possible ways to measure them. Further, we underscore the significance of fluctuating rheology to better understand micro structure - flow coupling. In sec. \ref{sec: Conclusions}, we synthesize our results (\ref{sec: Conclusions}).

\section{Background} \label{sec: Background}

\begin{figure*}
\centering
\begin{minipage}[b]{.35\textwidth}
\fcolorbox{red}{yellow}{\includegraphics[width = 1\textwidth]{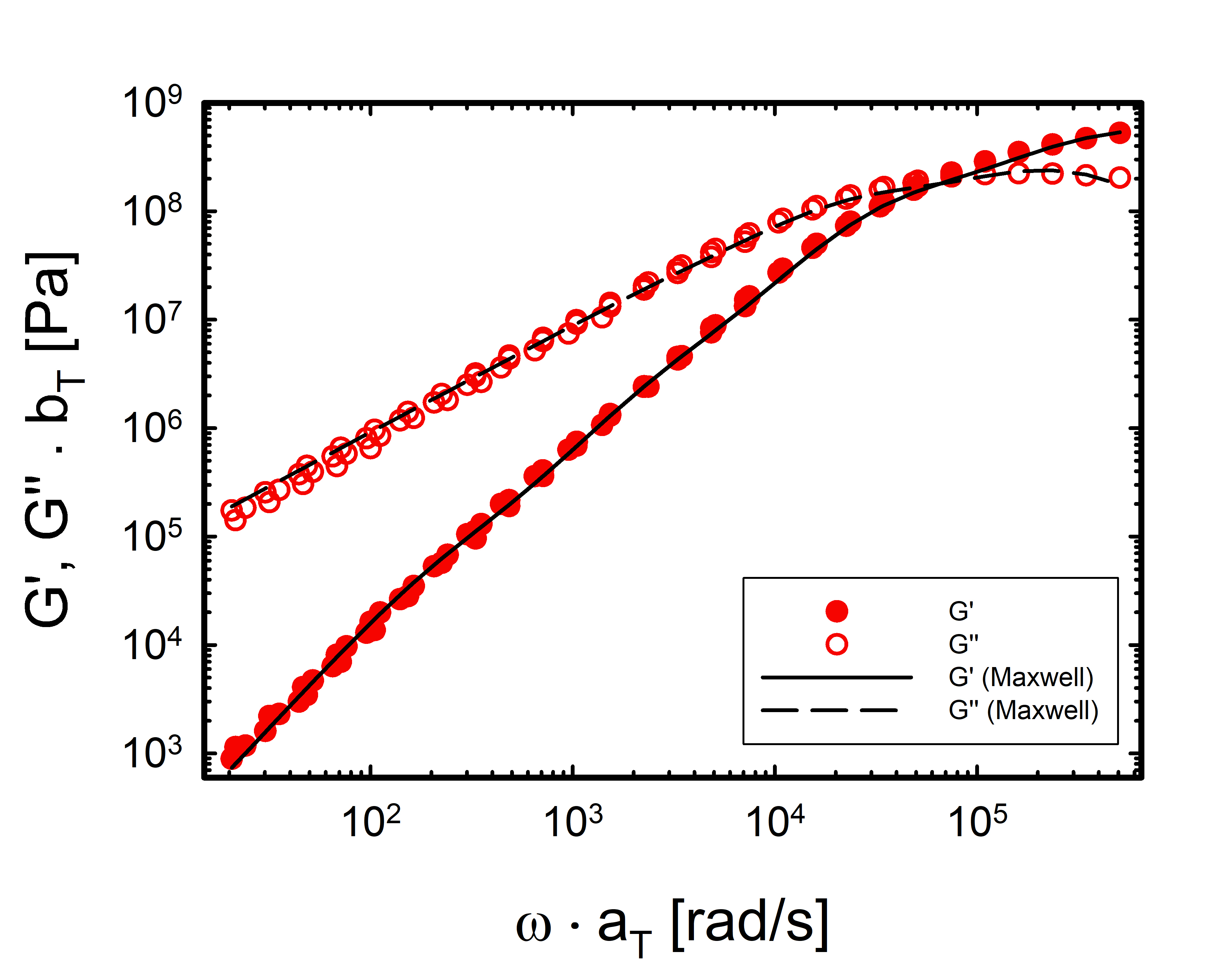}}
\caption{Linear viscoelastic moduli of an unentangled polymer melt of polystyrene, using time temperature superposition with $a_t$ the horizontal shift factor (data courtesy of \citet{Vincenzo}). A  bead-spring Rouse chain is used to fit the data, where the chain is composed of 4 connected dumbbells.}\label{fig: DB}
\end{minipage}\qquad
\begin{minipage}[b]{.35\textwidth}
\fcolorbox{blue}{yellow}{\includegraphics[width = 1.02\columnwidth]{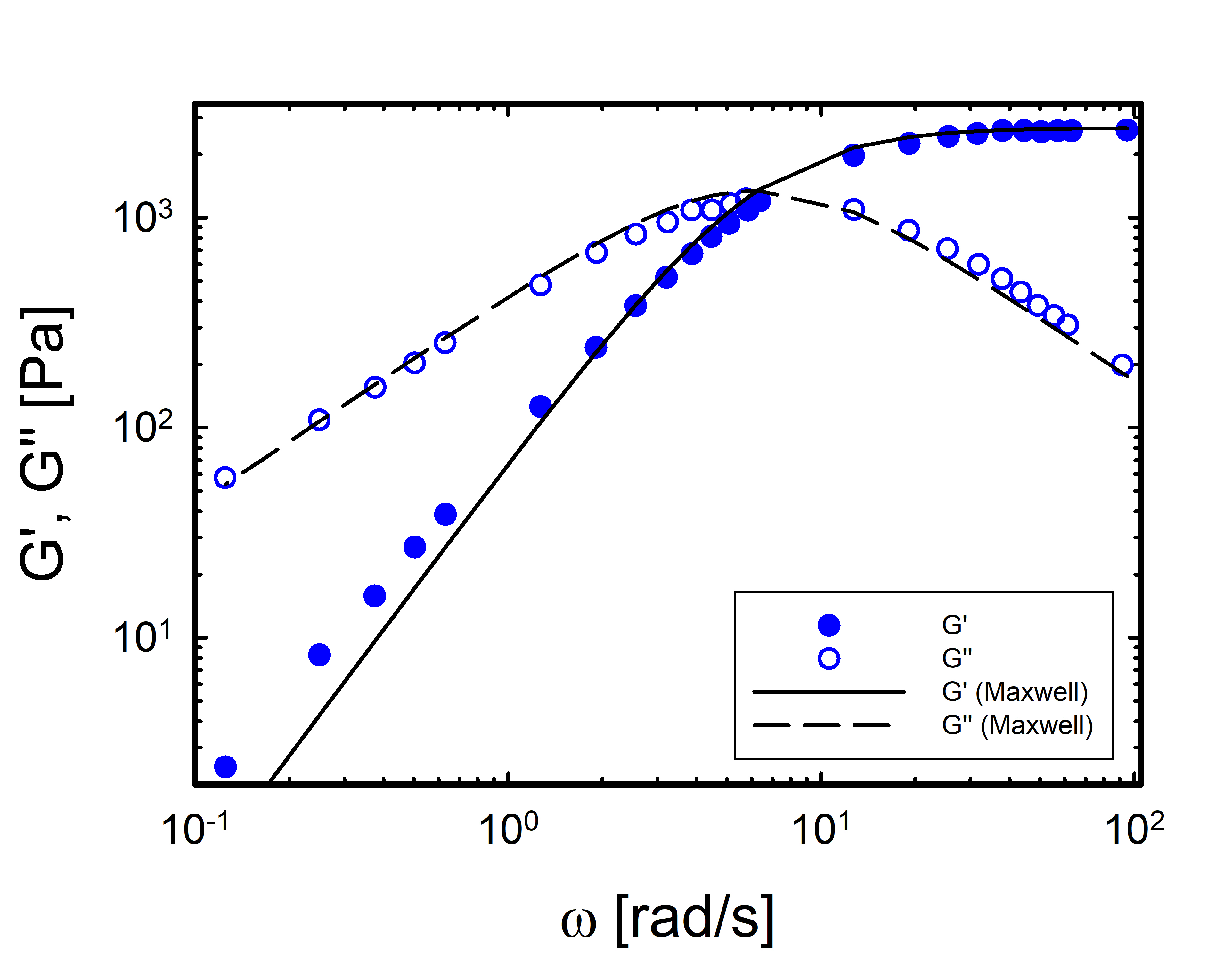}}
\caption{Linear viscoelastic moduli of associative telechelic polymers, replotted from Annable et al.\cite{ANNABLE} The data is fitted with a linear Maxwell model with one relaxation time. The dynamics of this physical polymer network can be represented by a temporary network model.}\label{fig: TNM_real}
\end{minipage}

\begin{minipage}[b]{.35\textwidth}
\fcolorbox{red}{yellow}{\includegraphics[width = 1.0\textwidth]{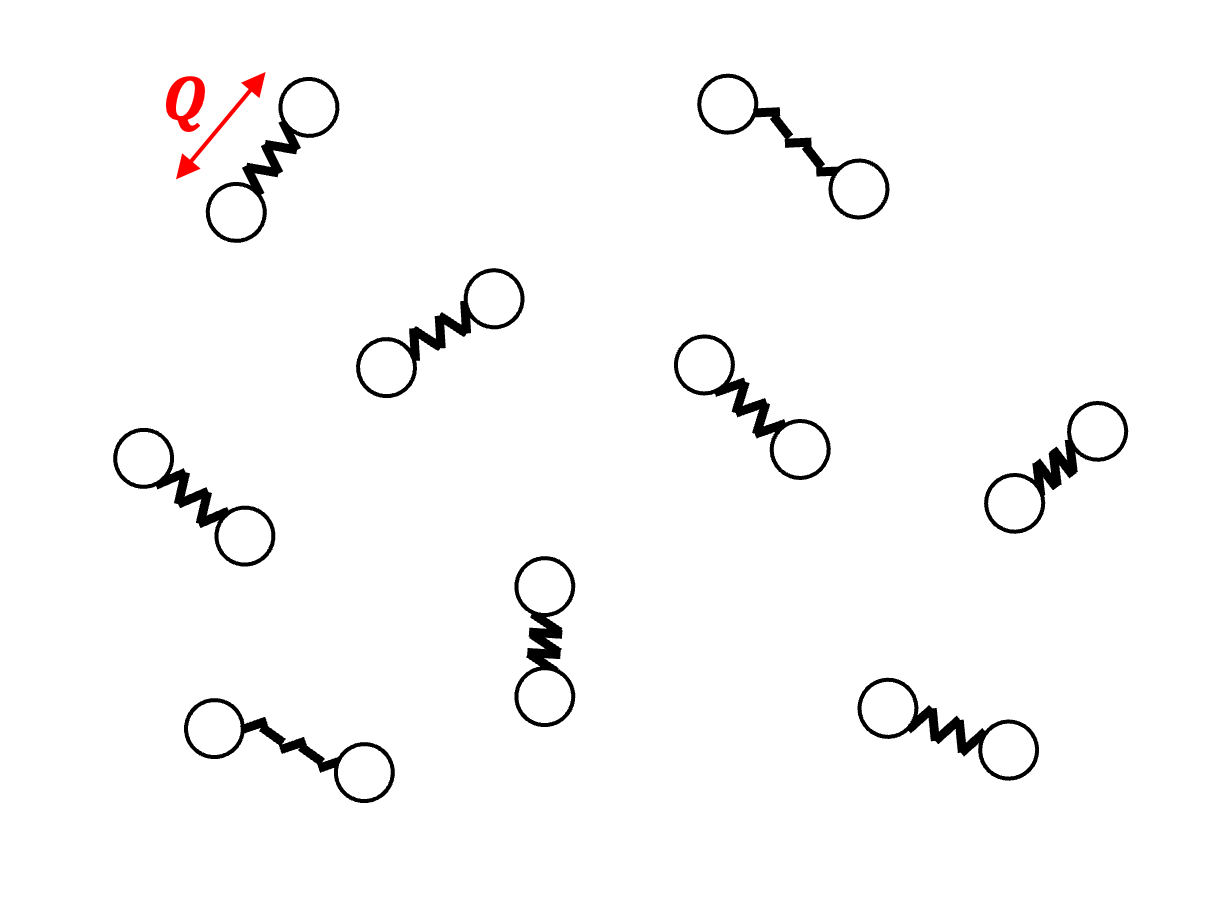}}
\caption{In the dumbbell model, dumbbells vibrate and move through the continuum volumes.}\label{fig: DB_Kramer}
\end{minipage}\qquad
\begin{minipage}[b]{.35\textwidth}
\fcolorbox{blue}{yellow}{\includegraphics[width = 1\textwidth]{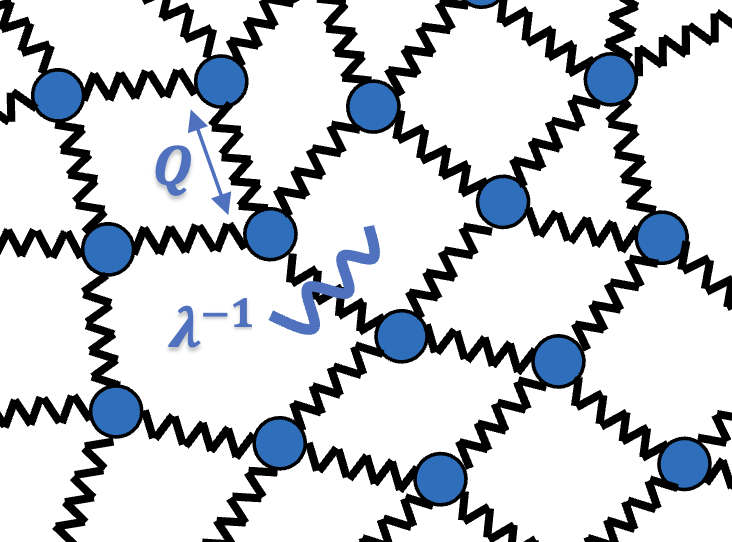}}
\caption{The temporary network model is composed of springs of a network that are destroyed (and generated) with rate $\lambda^{-1}$.}\label{fig: TNM}
\end{minipage}
\end{figure*}

\subsection{Temporary network and dumbbell model} \label{subsec: Models}
Maxwell introduced  the first, most simple and widespread phenomenological model for viscoelasticity in 1867, albeit for gases.\cite{Maxwell1866} Later it was enhanced by its observer-invariant formulation, known as the UCM,\cite{Oldroyd_50, Lodge_2} which relates the stress tensor $\pmb{\tau}$ to relative deformations:
\begin{equation} \label{eq: UCM}
 \partial_t{\pmb{\tau}}=\pmb{\kappa}\cdot\pmb{\tau}+\pmb{\tau}\cdot\pmb{\kappa}^T-G(\pmb{\kappa}+\pmb{\kappa}^T)-\frac{4H}{\zeta}\pmb{\tau},
\end{equation}
and its corresponding integral formulation, namely the \textit{Lodge equation}\cite{Lodge_2,Morrison2001}
\begin{equation}\label{eq: Lodge}
    \pmb{\tau}=\frac{G}{\lambda}\int_{-\infty}^t e^{-(t-t')/\lambda}\big[\pmb{\delta}-\pmb{C}^{-1}(t,t')\big]\d t'.
\end{equation}
$G$ is the modulus, $\lambda$ the relaxation time, $\pmb{\kappa}=(\nabla\pmb{u})^T$ is the transpose of the velocity gradient tensor and $\pmb{C}^{-1}$ the Finger strain tensor. These two models can be found in rheology textbooks \cite{Macosko, Larson, Transport2018, Phan_thien, Morrison2001,Ferry1980} and reproduce the rate and time dependency of the stress evolution of materials such as diluted polymer solutions (fig. \ref{fig: DB}) or networks of ssociative polymers with weak physical cross links (fig. \ref{fig: TNM_real}).

Two fundamentally different model systems, namely bead-spring dumbbells and temporary networks, effectively represent the microscopic dynamics associated with the macroscopic UCM.\cite{Rouse1953, Stochastic}
In the HDM, a springs with stiffness $H$ connects two beads at distance $\pmb{Q}$ (fig. \ref{fig: DB_Kramer}). Under the effect of a flow field the spring is stretched while the rapid interactions between the beads and the surrounding fluid are put into the model as Gaussian noise, as described in equation (eq. (\ref{eq: SDE})). In contrast, the TNM is rooted in rubber elasticity.\cite{Green_1946, Larson} The structural variable $\pmb{Q}$ represents the end-to-end vector between crosslinks (fig. \ref{fig: TNM}). Strands within this network deform in an affine manner and are subject to continuous cycles of breaking and reformation, occurring at a consistent rate $\lambda_{\text{TNM}}$. The introduction of new strands again ensures the adherence to a Gaussian equilibrium distribution $p_{\text{eq}}=(2\pi\sigma^2)^{-3/2}\exp{(-\pmb{Q}^2/\sigma^2)}$, with $\sigma^2$ being the characteristic equilibrium length.\\ \indent

These models possess three levels of description. The first level describes the evolution of $\pmb{Q}$.  The microscopic dynamics of HDM are described by a stochastic differential equation:
\begin{equation} \label{eq: SDE}
 \d\pmb{Q}=(\pmb{\kappa}\cdot\pmb{Q}-\frac{2H}{\zeta}\pmb{Q})\d t+\sqrt{\frac{4k_{\text{B}}T}{\zeta}}\d\pmb{W},
\end{equation}
where the $\cdot$ represent a contraction of indices, $\zeta$ the friction coefficient and $\pmb{W}$ is a vector, where each entry is an independent Wiener process.\cite{Stochastic}\\
The TNM starts from a deterministic equation for the advection,
\begin{equation} \label{eq: TNM convection}
\d\pmb{Q}=\pmb{\kappa}\cdot\pmb{Q}\d t,
\end{equation}
and one for the probability density of a strand's \enquote{lifetime} $t_{L}$\cite{Biller} 
\begin{equation} \label{eq: TNM destruction}
\phi(t_\text{L})=\frac{\exp{(-t_\text{L}\lambda_{\text{TNM}}^{-1})}}{\lambda_{\text{TNM}}}.
\end{equation}

Second, we seek the evolution of the probability density $p(t,\pmb{Q})$ (PDF). 
For the HDM we have\cite{Bird}
\begin{equation} \label{eq: Fokker}
 \partial_t p=-\partial_{\pmb{Q}}\cdot\left((\pmb{\kappa}\cdot\pmb{Q}-\frac{2H}{\zeta}\pmb{Q})p\right)+\frac{2k_{\text{B}}T}{\zeta}\partial_{\pmb{Q}}\cdot\partial_{\pmb{Q}}p,
\end{equation}
where $\partial_t,\partial_{\pmb{Q}}$ represent the partial differential operator with respect to time or the coordinates of the connector vector, respectively. For the TNM it is given by :
\begin{equation} \label{eq: prob_TNM}
\partial_t p+\partial_{\pmb{Q}}\cdot(\pmb{\kappa}\cdot\pmb{Q} p)=\lambda_{\text{TNM}}^{-1}[p_{\text{eq}}(\pmb{Q})-p(\pmb{Q},t)].
\end{equation}

Finally, to further coarse grain we extract the evolution equation of the 2\textsuperscript{nd} moment of $\pmb{Q}$, namely the conformation tensor $\pmb{c}(\pmb{x},t)=\langle\pmb{Q}\pmb{Q}\rangle=\int\pmb{Q}\pmb{Q}p \d^3 Q$, from (\ref{eq: Fokker}) and (\ref{eq: prob_TNM}). The latter, deterministic equation is quantitatively equal for both the TNM and the HDM for the identities $\lambda=\zeta/4H=\lambda_{\text{TNM}}$, $\sigma^2=k_{\text{B}}T/H$ and $G=nk_{\text{B}}T$, where $n$ represents the strand and dumbbell densities, respectively. Kramers' relation formalizes the momentum transmission over continuum volumes performed by the springs as follows:
\begin{equation} \label{eq: Kramer}
    \pmb{c}(\pmb{x},t)=nH\langle\pmb{Q}\pmb{Q}\rangle=nk_{\text{B}}T\pmb{\delta}-\pmb{\tau}(\pmb{x},t).
\end{equation}
This connection links the evolution equation for the configuration tensor to the stress and ultimately leads to the macroscopic version of the UCM (eq. (\ref{eq: UCM})), where $\pmb{c}$'s dependence on the location coordinate $\pmb{x}$ is due to $\pmb{\kappa}(\pmb{x},t)$. \\ \indent

A key question we seek to address is: At which level of description and up to which length scales do the two models differ?
This deeper dive into the HDM and the TNM shows that despite the different micro structural origin of the evolution of the PDF, the second order moments evolve in the same manner, so that the stress follows the UCM (or the integral version , the Lodge model). However, a closer look at the stress fluctuations up to some lengthscale will reveal the microscopic differences of the models.  This requires us however, to first introduce a suitable formulation of the FDT. 

\subsection{Relating fluctuations and dissipation} \label{subsec: Fluctuation dissipation}
The FDT, first formulated in 1928 by \citet{Nyquist1928}, creates a connection across length scales. It links the behavior of fluctuating microscopic world or models to the dissipative part of associated macroscopic equations.\\ \indent

Models based on the FDT can capture important properties of complex fluids even in non-equilibrium situations. For instance, the occurrence  of $\zeta$ in both  terms for the noise and the deterministic motion (eq. (\ref{eq: SDE})) mirrors the FDT in Rouse bead-spring chains. Current research efforts\cite{Ianniruberto19, Ianniruberto20} rely on the FDT to replicate strongly nonlinear properties of polymer melts. Recently, the FTD has been extended to active matter.\cite{Brady2019}\\ \indent

A rigorous derivation of the FDT is restricted to thermodynamic equilibrium, limiting its direct applicability to scenarios involving only linear perturbations. This is also known as the FDT of the first kind. Conversely, the FDT of the second kind extends to encompass more complex dynamics, such as those observed in e.g. strong shear flows, where traditional linear approaches fall short. It consists in linking the dissipative structure of the macroscopic evolution to the distribution of their fluctuations.\cite{Alberto} The credibility of this extended application is supported by empirical successes and theoretical justifications based on fundamental principles.\cite{Beyond} Yet its validity remains an open debate (e.g. \citet{Watanabe2023}).
\citet{Donev} approximated reaction-diffusion systems with the Landau-Lifshitz approach (eq. (\ref{eq: Fluct_hydrodynamics})) and found inconsistencies in the results for small ensembles of molecules. A successive work by Montefusco et al.\cite{Alberto} explained the latter problems by revealing how not all processes can be consistently described in terms of stochastic differential equations, since non-Gaussian noise requires a different description.\\ \indent 

In this paper we show that also the TNM requires a description beyond of non-Gaussian noise. To facilitate this investigation, we introduce the framework of GENERIC (General Equation for Non-Equilibrium Reversible-Irreversible Coupling) formalism.\cite{Grmela_Oe1,Grmela_Oe2,Beyond} This approach uniquely combines reversible dynamics with an irreversible dissipative component, providing a cohesive structure for articulating the FDT in a very natural manner. GENERIC expresses the evolution of independent variables $\pmb{x}$ in an  nonequilibrium system isolated from its surroundings, through the total energy $E$, the total entropy $S$ and two linear operators $\pmb{L},\, \pmb{M}$, which capture the physical constraints
\begin{equation} \label{eq: GENERIC 1}
 \frac{\d \pmb{x}}{\d t}=\pmb{L}(\pmb{x})\cdot\partial_{\pmb{x}}E(\pmb{x})+\pmb{M}(\pmb{x})\cdot\partial_{\pmb{x}} S(\pmb{x}).
\end{equation}
As our focus lies on the dissipative characteristics, we omit the reversible contribution ($L=0$) for the remainder of this work. The fluctuations link to the irreversible component of eq. (\ref{eq: GENERIC 1}) through the FDT in its widespread Green-Kubo formulation\cite{Kubo,Beyond}
\begin{equation} \label{eq: Green-Cubo}
 \pmb{B}(\pmb{x})\cdot \pmb{B}(\pmb{x})=2k_{\text{B}}\pmb{M}(\pmb{x}),
\end{equation}
where $\pmb{B}(\pmb{x})$ describes the Gaussian fluctuations with the stochastic differential equation for the  stochastic process $\pmb{X}_\tau$ given by:
\begin{equation} \label{eq: Diffusive SDE}
 \d \pmb{X}_t = \bigr[ \pmb{M}(\pmb{x})\cdot\partial_{\pmb{x}}S(\pmb{x})+k_{\text{B}}\partial_{\pmb{x}}\cdot \pmb{M}(\pmb{x})\bigr]_{\pmb{x}=\pmb{X}_t} \d t + \pmb{B}(\pmb{X}_t)\cdot \d\pmb{W}_t.
\end{equation}
The necessity for this formulation arises from its foundational role in modeling fluctuations within complex fluids. Of particular relevance to our paper is the elucidation of noise enhancement dynamics. Specifically, for Gaussian noise, one can readily isolate the dissipative component of the macroscopic evolution equation given by the friction matrix $\pmb{M}$, as depicted in eq. \ref{eq: GENERIC 1}. Subsequently, the noise can be readily enhanced on the system with the formulation eq. (\ref{eq: Diffusive SDE}). Conversely, given the noise component ($\pmb{B}$), one can infer the corresponding macroscopic equation $\pmb{M}$. To date, this relationship has been perceived as a comprehensive means of encapsulating noise in rheology.\cite{Huetter2020, Beyond}

\indent
A second formulation withing GENERIC uses the dissipation potential $\Psi^*(\pmb{x},\pmb{\xi})$, instead of $\pmb{M}$
\begin{equation} \label{eq: GENERIC 2}
 \frac{\d \pmb{x}}{\d t}=\partial_{\pmb{\xi}}\psi^*(\pmb{x},\pmb{\xi})\big|_{\pmb{\xi}=\partial_{\pmb{x}} S(\pmb{x})}.
\end{equation}
Among the differences between eq. (\ref{eq: GENERIC 1}) and (\ref{eq: GENERIC 2}),\cite{Huetter2013} we highlight that eq. (\ref{eq: GENERIC 1})-(\ref{eq: Green-Cubo}) is limited to to the description of Gaussian noise, while $\psi^*$ can be used to model general Markow processes according to a generalized FDT\cite{Oettinger_1, Alberto,Alberto_thesis}
\begin{equation} \label{eq: General FDT}
 G_\text{C}=\frac{2k_{\text{B}}}{\tau}\ln{\langle e^{\alpha(\pmb{X}_\tau-\pmb{x})}\rangle}\approx\psi^*(\pmb{x},\pmb{\xi})-\psi^*(\pmb{x}, \partial_{\pmb{x}}S(\pmb{x}))
\end{equation}
with $G_\text{C}$ as the cumulant generating function of the stochastic process $\pmb{X}_\tau$ in a time interval $\tau$. In presence of fluctuations for $\alpha$ we have $ \xi - \frac{\delta S}{\delta p} = 2 k_{\text{B}} \alpha$. Consequently, we can think of $\xi$ as representing the distance of the system form equilibrium. When $\xi=0$, the potential $\psi^*$ equals its minimum at zero i.e. there is no dissipation in the evolution of the macroscopic system. Eq. (\ref{eq: General FDT}) captures how with growing distance from equilibrium (larger values of $\xi$) the dissipative part of the macroscopic evolution equation changes. 

\section{Dissipation potentials} \label{sec: Potentials}
\subsection{Origin of $\pmb{c}$ fluctuations in the two models} \label{subsec: Origins}
\begin{figure} 
\centering
\includegraphics[width=0.48\textwidth]{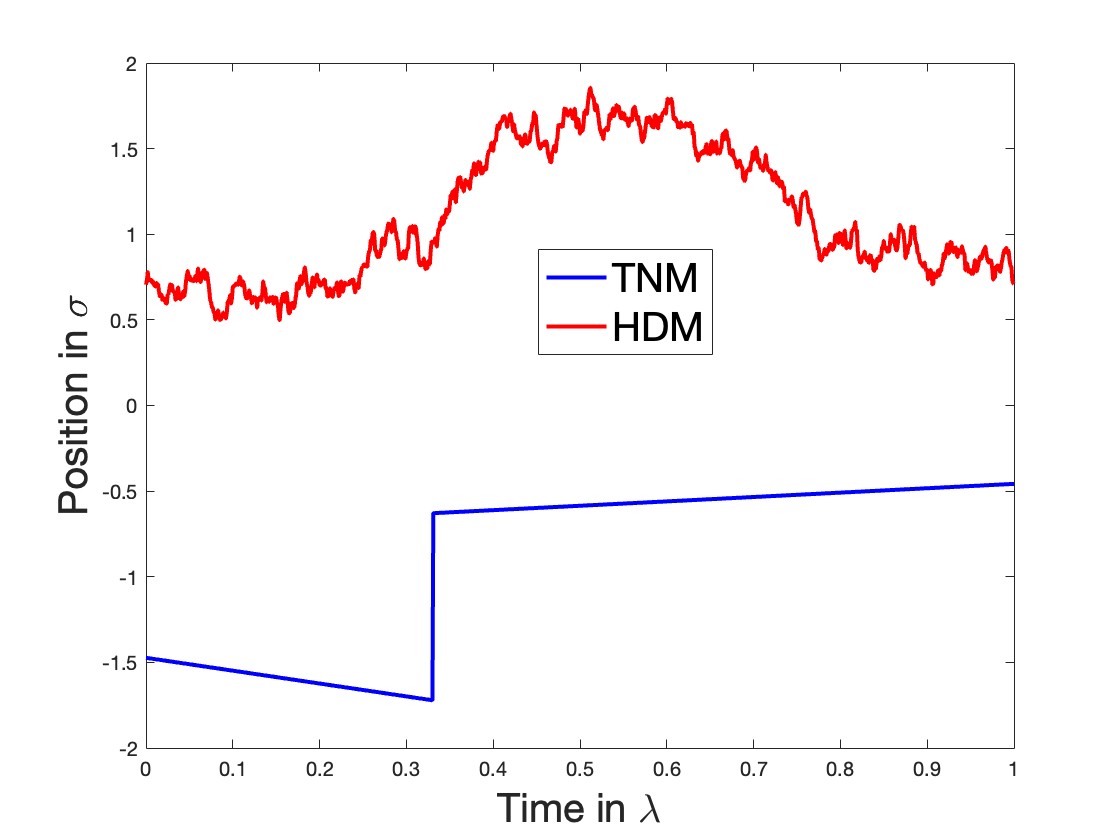}
 \caption{Comparison of the individual trajectories in a simple shear flow of $\pmb{Q}$'s component in flow direction ($Q_1$). For the dumbbell, the frequent, small collisions of the beads with the surrounding microscopic particles enable the use of continuous sample paths with superimposed Gaussian white noise $\d\pmb{W}_t$ (eq. (\ref{eq: SDE})). In contrast, the temporary network model exhibits deterministic trajectories, abruptly interrupted by stochastic jumps when strands are replaced.} \label{fig: Trajctories}
\end{figure}

\begin{figure}%
\centering
\includegraphics[width=0.5\textwidth]{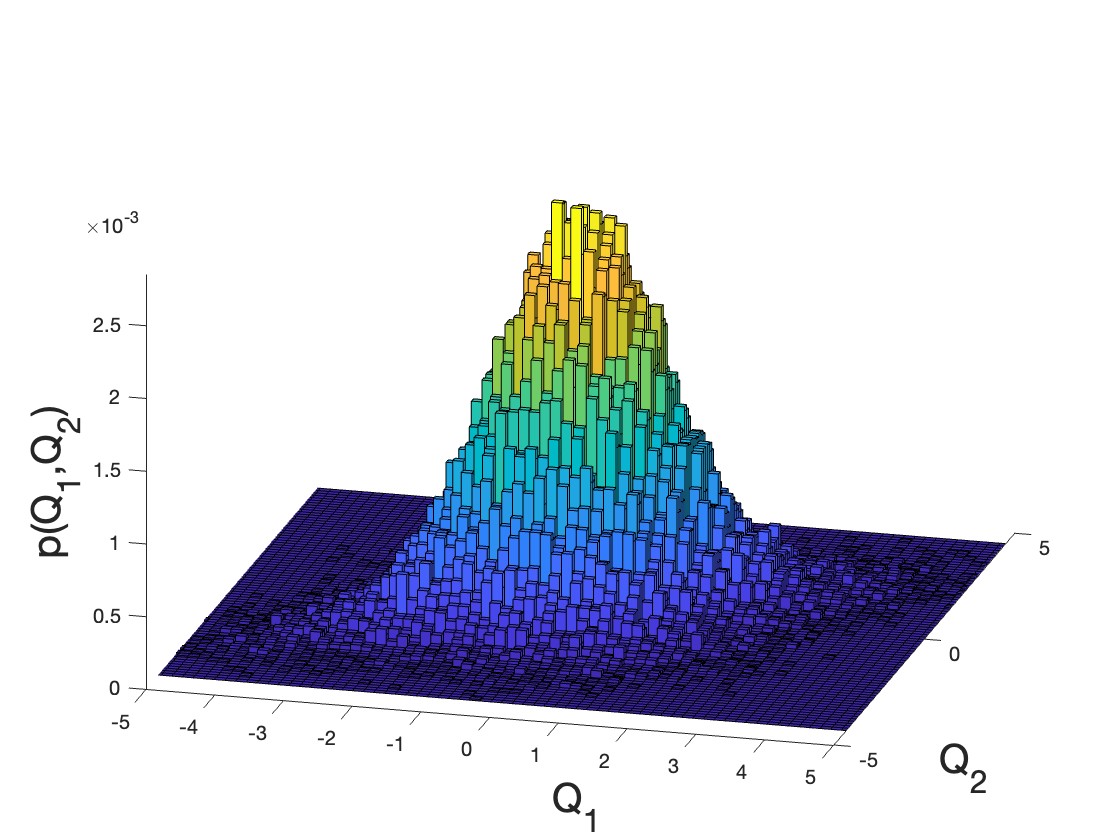}
 \caption{Typical histogram of the PDF for the temporary network model during a step strain experiment at $t=\lambda\ln{2}$. Strands \enquote{jump} from one -column $V_j$ to another when  they are replaced. This stochastic changes in the histogram cause subsequent fluctuations in the materials fluxes.} 
 \label{fig: Histogram}%
\end{figure}

To understand the difference in the fluctuations of the two models we first focus on single trajectories of $\pmb{Q}$'s components, for example in flow direction ($Q_1$) (eq. (\ref{eq: SDE}) and (\ref{eq: TNM convection}-\ref{eq: TNM destruction})). Fig. \ref{fig: Trajctories} shows the different origin of fluctuations in $\pmb{c}$, namely the trajectories of $\pmb{Q}$.
The latter and the following figure results from our simulations. We solve the linear stochastich differential equation (\ref{eq: SDE}) of the HDM with the Euler-Maruyama scheme.\cite{Stochastic} For the TNM we apply the algorithm presented by \citet{Biller} to solve (\ref{eq: TNM convection})-(\ref{eq: TNM destruction}). Basically the stochastic element in the trajectory is introduced through a Gillespie algorithm\cite{Gillespie} for the strand's lifetime. 
The simulations are validated by comparison to analytical solution for $\pmb{c},p(\pmb{Q})$. 
In the HDM noise is an intrinsic part of the dynamics (fig. \ref{fig: Trajctories}). The frequent, small interactions of microscopic particles allow the usage of continuous sample paths with superposed Gaussian white noise $\d\pmb{W}_t$. In contrast, the TNM exhibits deterministic trajectories abruptly interrupted by stochastic jumps when strands are replaced. Noise becomes apparent only at the coarse-grained level of a discretized PDF (\ref{eq: prob_TNM}), where the trajectories of a finite number of strands are aggregated into a histogram. If the space of possible $\pmb{Q}$ strands is partitioned in cells $j$ with volume $V_j$, the relative frequency of ensemble members in the volume $V_j$ defines the probability $P_j$. The piecewise constant function with value $p_j=P_j/V_j$ on the cell $j$ provides a noisy histogram approximation to the probability density $p(\bm{Q})$ (see fig. \ref{fig: Histogram}). Thus, the strand's \enquote{jumps} cause fluctuations in the columns of the histogram.

\subsection{The dumbbells}
Since the noise of the dumbbells  is uncorrelated, the FDT results in the well known Green-Kubo relation (\ref{eq: Green-Cubo}), which has been exploited to formulate noise in $\pmb{c}$ in past works.\cite{Huetter2020,Beyond} To express the PDF in terms of GENERIC and thus of $p(\pmb{Q})$ functionals we have\cite{Grmela_Oe2,Beyond}

\begin{equation} \label{eq: M_44 pot}
\begin{split}
 & S[p] = \int-p\frac{H\pmb{Q}^2}{2T}-pk_{\text{B}}\ln{p}\d Q^3, \\
 & M[p] = -\partial_{\pmb{Q}}\cdot\frac{2T}{\zeta}p(\pmb{Q})\partial_{\pmb{Q}}[\cdot]. \\
\end{split}
\end{equation}
It is then possible to construct a quadratic potential $\psi^*$
\begin{equation}\label{eq: Psi_DB}
    \psi^*[p,\xi]=\frac{1}{2}M(\pmb{Q})\xi^2=-\frac{1}{2}\partial_{\pmb{Q}}\cdot\frac{2T}{\zeta}p(\pmb{Q})\partial_{\pmb{Q}}[\xi^2],
\end{equation}
so that both eq. (\ref{eq: GENERIC 1}) and eq. (\ref{eq: GENERIC 2}) reproduce the dissipative part of eq. (\ref{eq: prob_TNM}):
\begin{equation}\label{eq: irr_DM_PDF}
\begin{split}
    &\frac{\d p}{\d t}=\partial_{\pmb{\xi}}\psi^*(\pmb{x},\pmb{\xi})\big|_{\pmb{\xi}=\partial_{\pmb{x}} S(\pmb{x})}\\
    &=\partial_{\pmb{Q}}\cdot\Big[\frac{2T}{\zeta}p\partial_{\pmb{Q}}[\frac{H\pmb{Q}^2}{2T}+k_{\text{B}}\ln{p}]\Big]\\
    &=\partial_{\pmb{Q}}\cdot(\frac{2H}{\zeta}\pmb{Q}p)+\frac{2k_{\text{B}}T}{\zeta}\partial_{\pmb{Q}}\cdot\partial_{\pmb{Q}}p.
    \end{split}
\end{equation}
Now, that eq. (\ref{eq: Psi_DB}) allows us to express the irreversible part of the PDF (\ref{eq: irr_DM_PDF}), we can derive $\psi^*$ for the TNM to compare the two dissipation potentials.

\subsection{The temporary network} \label{subsec: TNM fluctuations}
One may verify that an entropy
\begin{equation}\label{eq: Entropy}
 S[p]=-k_{\text{B}}\int p(\pmb{Q})\ln{\frac{p(\pmb{Q})}{p_{\text{eq}}(\pmb{Q})}}\d^3 Q,
\end{equation}
and a potential 
\begin{equation} \label{eq: Psi_TNM}
\psi^*[p,\xi]=\frac{4k_{\text{B}}}{\lambda}\int\sqrt{p(\pmb{Q})p_{\text{eq}}(\pmb{Q})}\big(\cosh{\frac{\xi(\pmb{Q})}{2k_{\text{B}}}-1}\big)\d^3Q
\end{equation}
reproduce the right-hand side of (\ref{eq: prob_TNM}).\\

Now, armed with an enhanced understanding of fluctuations, as detailed in sec. \ref{subsec: Origins} and illustrated in fig. \ref{fig: Histogram}, we proceed to apply the FDT as formulated in eq. (\ref{eq: General FDT}). This application involves a two-step limiting process - initially expanding the ensemble sizes towards infinity, followed by the reduction of the volumes $V_j$ of the cells of the observation cells to infinitesimally small dimensions. 
Through this procedure, the resulting noisy histograms gradually approximate the probability distribution  $p(\pmb{Q})$. The theory of large deviations provides a quantitative framework for describing this convergence,\cite{Alberto_thesis} offering precise insights into the behavior of these distributions. In the realm of mathematics, this concept finds a more rigorous formulation through the use of empirical measures. Here, we use the physical insight of the transition from discrete $P_j$ to $p(\pmb{Q},t)$ to write
\begin{equation}\label{eq: Discrete}
 \int \alpha_N(\bm{Q}) \Delta p(\bm{Q}) d^3Q = \sum_j \alpha_N(\bm{Q}_j) \Delta P_j ,
\end{equation}
where $\pmb{Q}_j \in V_j$ and $\alpha_N(\bm{Q})$ indicates an auxiliary quantity dependent on the $N$ strands constituting the ensemble.
The symbol $\Delta$ indicates the stochastic change in the time lapse $\tau$. For small $\tau/\lambda$ we approximate the process by a discrete-time Markov jump process. The transitions are the results of two independent Poisson processes $L^+_j,\,L^-_j$, where we justify the independence with two assertions. First, $L^+_j$ depends on the entire ensemble, while $L^-_j$ only depends on the members in cell $j$. Secondly, the constraint of a constant ensemble size has no significant effect on the independence property. A change in the ensemble composition can either take place, because strands \enquote{die} with average $p_{-}=\frac{\tau}{\lambda}V_jp_jN$, or because they form with average $p_{+}=\frac{\tau}{\lambda}V_jp_{\text{eq}}N$.\\
Given the non-Gaussian nature of the noise, we cannot use relation (\ref{eq: Green-Cubo}). However, with the discrete representation of the stochastic process for small ensembles, we express the left hand side of the generalized FDT (\ref{eq: General FDT})
\begin{equation} \label{eq: Kum_Gen_TNM}
\begin{split}
 &G_C =\frac{2k_{\text{B}}}{\tau}\ln{\big\langle\exp{\sum_j\frac{\alpha_N(\pmb{Q}_j)}{N}(L_j^{+}-L_j^{-})}\big\rangle}\\
 &\overset{\text{ind.}}{=}\frac{2k_{\text{B}}}{\tau}\sum_j\ln{\big\langle\exp{\frac{\alpha_N(\pmb{Q}_j)}{N}L_j^{+}}\big\rangle}+\ln{\big\langle\exp{-\frac{\alpha_N(\pmb{Q}_j)}{N}L_j^{-}}\big\rangle}\\
 &\overset{\text{Poi.}}{=}\frac{2k_{\text{B}} N}{\lambda}\sum_j V_j \big[p_{\text{eq}}(\pmb{Q}_j)(e^{\alpha_N(\pmb{Q}_j)/N}-1)+p(\pmb{Q}_j)e^{\alpha_N(\pmb{Q}_j)/N}-1) \big]\\
 &=\frac{2k_{\text{B}} N}{\lambda}\Big(\int p_{\text{eq}}(\pmb{Q})e^{\alpha_N(\pmb{Q})/N}+p(\pmb{Q})e^{-\alpha_N(\pmb{Q})/N}\d^3Q-2 \Big).
\end{split}
\end{equation}
In the second line we used the independence of the two processes, in the third the cumulant generating function of a Poisson random variable, in the fourth we assumed the limit of large ensembles and vanishing $V_j$. 
$G_C$ is independent of the time interval $\tau$ and of the particular discretization of the PDF. However, it depends on the ensemble size $N$. Large deviation theory considers the effect of a single jump event, so that we are interested in $N=1$ and write $\alpha_1(\bm{Q})=\alpha(\bm{Q})$. Then, the general FDT (\ref{eq: General FDT}) holds as $\psi^*$ from eq. (\ref{eq: prob_TNM}) agrees with the microscopic fluctuations in eq. (\ref{eq: Kum_Gen_TNM}).\\
This result reflects the distinct jump dynamics for one single strand depicted in fig. \ref{fig: Trajctories}. The implication for the fluctuations in $\pmb{c}$ and thus $\pmb{\tau}$ are further explored in sec. \ref{sec: Fluctuations comparison}. The potential (\ref{eq: Psi_TNM}) resembles more the one of chemical reactions\cite{Alberto} than the one of diffusion processes (\ref{eq: Psi_DB}). This is no surprise as the dynamic of the processes is close to the \enquote{jumps} of a chemical reaction.
An expansion of the $\cosh{}$ function shows that at higher values of the non-equilibrium parameter $\xi$ the potential ($\ref{eq: Psi_TNM}$) differs more and more from the quadratic potential of diffusive process as of the HDM.

\section{Differences in the PDFs} \label{sec: PDF}
At equilibrium both HDM and TNM exhibit the same distribution $p_{\text{eq}}$. Under flow the stress $\pmb{\tau}$ in  both models is described by the UCM (which also holds for the  evolution of $\pmb{c}$. However, sec. \ref{sec: Potentials} implies different fluctuations in $\pmb{c}$ and thus underlying different PDFs. To characterize this we calculate the evolution equation of the  4\textsuperscript{th} order moment, which for the Gaussian HDM model evolves according to Wick's theorem :
\begin{align}
\begin{split} \label{eq: 4th moment DB}
\partial_t \mu_{abcd}&=\frac{\partial u_a}{\partial x_j}\mu_{jbcd}+\mu_{ajcd}\frac{\partial u_b}{\partial x_j}+\mu_{abjd}\frac{\partial u_c}{\partial x_j}+\mu_{abcj}\frac{\partial u_d}{\partial x_j}\\ 
&-\frac{8H}{\zeta}\mu_{abcd}+\frac{4k_{\text{B}}T}{\zeta}(\delta_{ab}\mu_{cd}+\delta_{bc}\mu_{ad}+\mu_{ab}\\ &\delta_{cd}+\delta_{ac}\mu_{bd}+\delta_{bd}\mu_{ac}+\delta_{ad}\mu_{bc}),
 \end{split}
\end{align}
For the non-Gaussian TNM PDF the  4\textsuperscript{th} order moment evolves accoridng to:
\begin{align}
 \begin{split} \label{eq: 4th moment TNM}
 \partial_t \mu_{abcd}&=\frac{\partial u_a}{\partial x_j}\mu_{jbcd}+\mu_{ajcd}\frac{\partial u_b}{\partial x_j}+\mu_{abjd}\frac{\partial u_c}{\partial x_j}+\mu_{abcj}\frac{\partial u_d}{\partial x_j}\\
 &+\lambda^{-1}(\mu_{abcd}^{\text{eq}}-\mu_{abcd}),
 \end{split}
\end{align}
where $\mu_{abcd}^{\text{eq}}$ is the equilibrium 4\textsuperscript{th} moment.\\

\begin{figure}%
\centering
\includegraphics[width=0.5\textwidth]{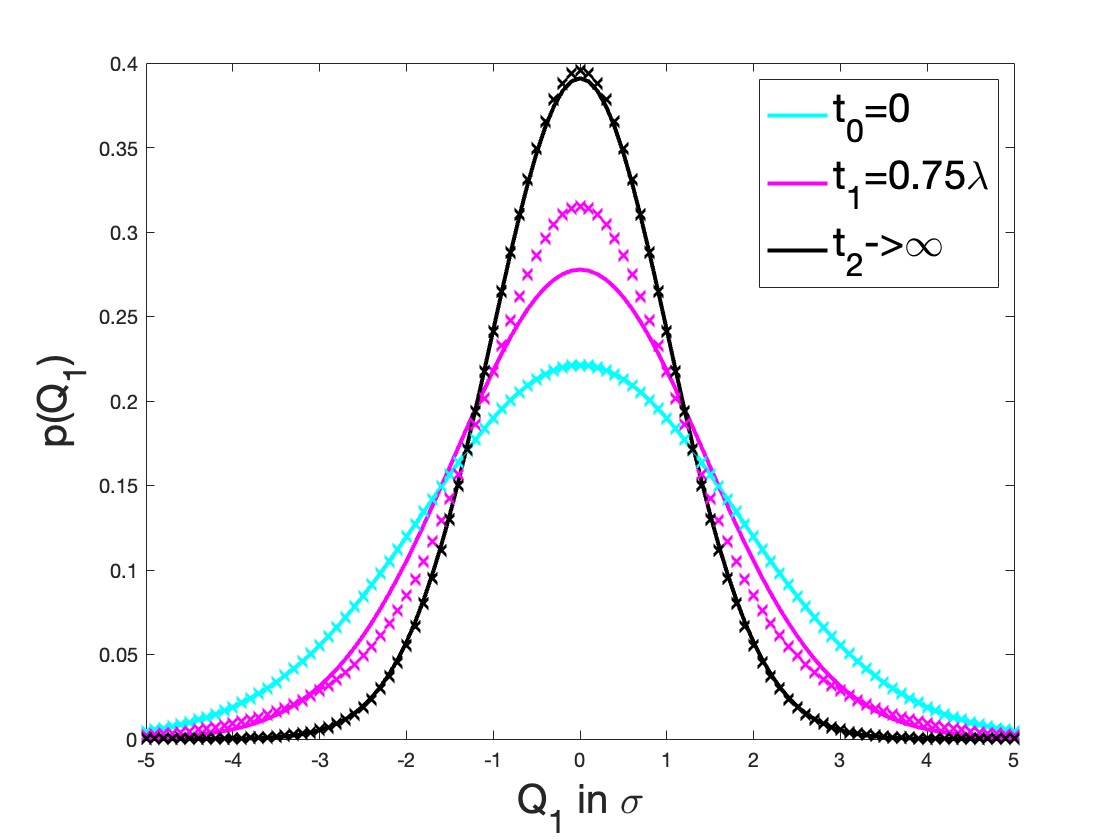}
 \caption{The probability density function of the temporary network (crosses) and the dumbbells (line) for different times after applying an instantaneous step strain at $t=0$. The strain is of $1.5$ instead of $1$ to accentuate the difference.} 
 \label{fig: PDFs}%
\end{figure}

\begin{figure}
\centering
\includegraphics[width=0.5\textwidth]{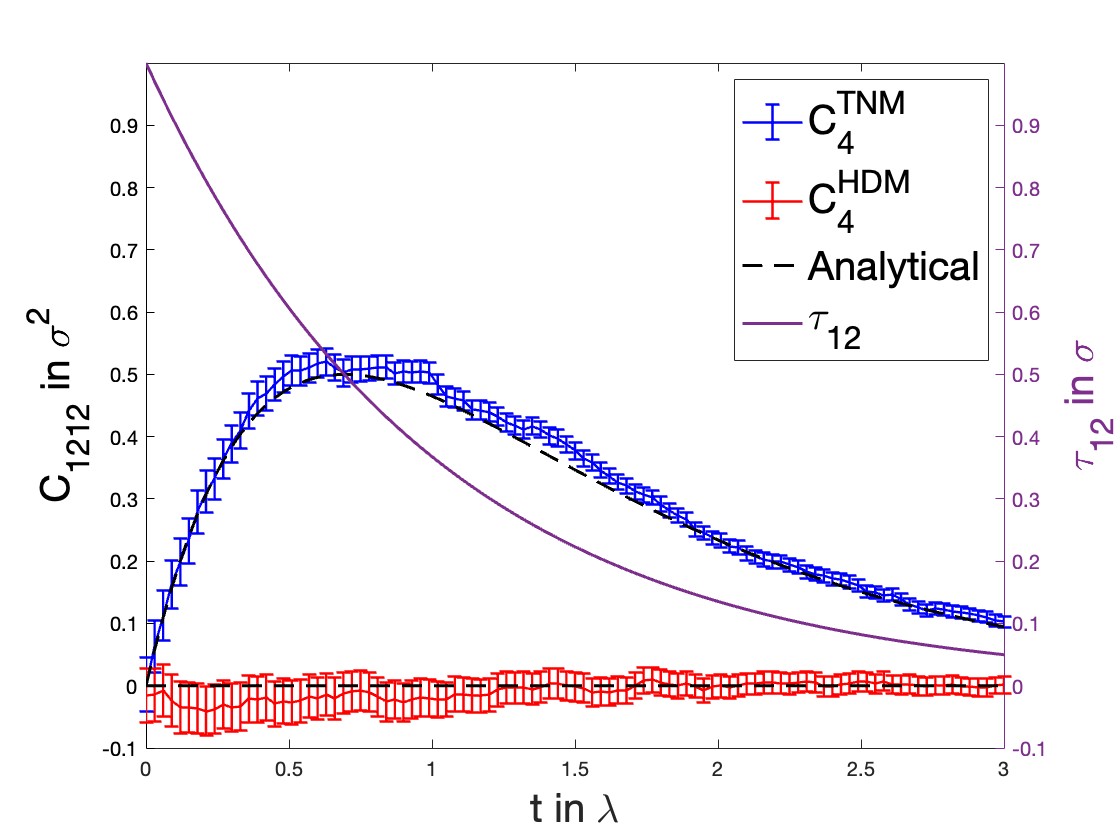}
 \caption{4\textsuperscript{th} cumulant $C_{1212}$ as for a step strain ($\gamma=1$), an ensemble of $N=10^5$ and a time discretization of $\d t=0.03\lambda$ for both the temporary network and the dumbbells. The maximum in the difference occurs at $t=\lambda\ln{2}$ and relaxes slowly - in fact only for $t\rightarrow\infty$ the difference ceases to exist.} 
 \label{fig: Cumulants}%
\end{figure}

\begin{figure}%
\centering
\includegraphics[width=0.5\textwidth]{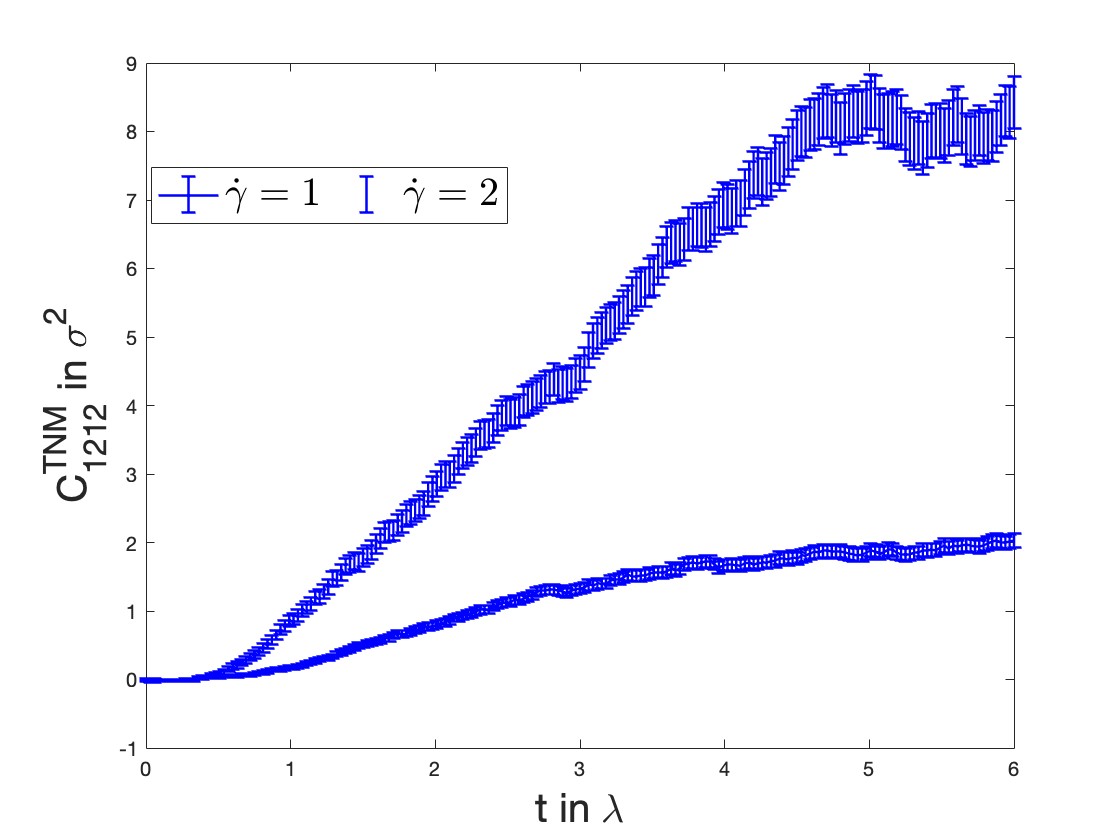}
 \caption{4\textsuperscript{th} cumulant $C_{1212}$ of the TNM for start up flows at two different shear rates, an ensemble of $N=10^5$ and a time discretization of $\d t=0.03\lambda$. In agreement with the findings of sec. \ref{sec: Potentials}, the further the systems are from equilibrium (larger values of $\xi$, or specifically for this example $\dot{\gamma}$), the more the TNM and the HDM differ from eachother.} 
 \label{fig: Start_up_flow}%
\end{figure}

Analytical solutions to the step strain experiment can be used to illustrate the differences between the evolution of the PDF's. Before the instantaneous deformation is imposed both model predict $p_{\text{eq}}$ at equilibrium. The shear rate equals $\Dot{\gamma}=\gamma\delta(t)$, so that integration of (\ref{eq: SDE}) and (\ref{eq: TNM convection})-(\ref{eq: TNM destruction}) from $0^-$ to $0^+$ leads to $\pmb{Q}_0=\Gamma\cdot\pmb{Q}_{\text{eq}}$ with $\pmb{\Gamma}=\exp{(\nabla\pmb{u}^T)}$, where the subscript eq refers to equilibrium opposed to $0$ as the initial condition. Since the configurations are convected, the PDFs remain Gaussian with first and second order moment $\langle Q_1 Q_1\rangle$ is proportional to the normal stress $\tau_{11}$ according to 
\begin{align} \label{eq: Initial cond}
    \langle\pmb{Q}\rangle_0=\pmb{\Gamma}\cdot\langle\pmb{Q}\rangle_{\text{eq}},\\
    \langle\pmb{Q}\pmb{Q}\rangle_0=\pmb{\Gamma}\cdot\langle\pmb{Q}\pmb{Q}\rangle_{\text{eq}}\cdot\pmb{\Gamma}^T.
\end{align}
To simplify the graphical representation we restrict ourselves to the $Q_1$ component so to observe the marginal distribution function $p(Q_1)=\int\int p(\pmb{Q})\d Q_2 \d Q_3$. Then eq. (\ref{eq: prob_TNM}) results in the superposition of two contributions
\begin{equation} \label{eq: Step_strain_TNM_p}
    p_{\text{TNM}}(t)=p_{\text{eq}}(1-\exp{(-\lambda^{-1}t)})+p_0\exp{(-\lambda^{-1}t)}.
\end{equation}
On the other hand, the PDF evolution eq. (\ref{eq: Fokker}) is a Fokker-Planck equation with linear coefficients. It consequently leads to a Gaussian distribution uniquely determined by its first and second moments with known respective evolution equations.\cite{Transport2018} Fig. \ref{fig: PDFs} compares the two different analytical solutions. Only at equilibrium (i.d. at $t<0$ or $t\rightarrow\infty$) and at the initial condition $t=0$ are the two PDFs equal. At all intermediate times, $p_{\text{TNM}}$ is non-Gaussian. The distinction is best captured by fig. \ref{fig: Cumulants}. It depicts the fourth cumulant $C_{1111}$, which in 1-D corresponds to $C_{1111}=\mu_{1111}-3\mu_{11}^2$ and thus reflects the differences in eq. (\ref{eq: 4th moment DB}) and (\ref{eq: 4th moment TNM}).\\
For flows strengths beyond our analytical framework the discrepancy is visible in numerical simulations. For instance, in fig. \ref{fig: Start_up_flow}, the impact of different shear rates ($\dot{\gamma}$) on a start-up flow is depicted in relation to the differences $\mu_{1212}^{\text{TNM}}-\mu_{1212}^{\text{HDM}}$. Given that the distribution for the HDM is always Gaussian, this difference equates $C_{1212}^{\text{TNM}}$, since $C_{1212}^{\text{HDM}}$ is constant $0$. In agreement with the findings of sec. \ref{sec: Potentials}, the stronger the perturbation from equilibrium, the larger the difference.\\
Until this point the discussion was confined to the configurational space of $\pmb{Q}$, its single trajectories or PDFs. In the following section we elucidate how these result relate to the stress fluctuations in actual volumes.

\section{The limit of Gaussian noise for stress fluctuations in complex fluids} \label{sec: Fluctuations comparison}
\begin{figure}
\centering
\includegraphics[width=0.5\textwidth]{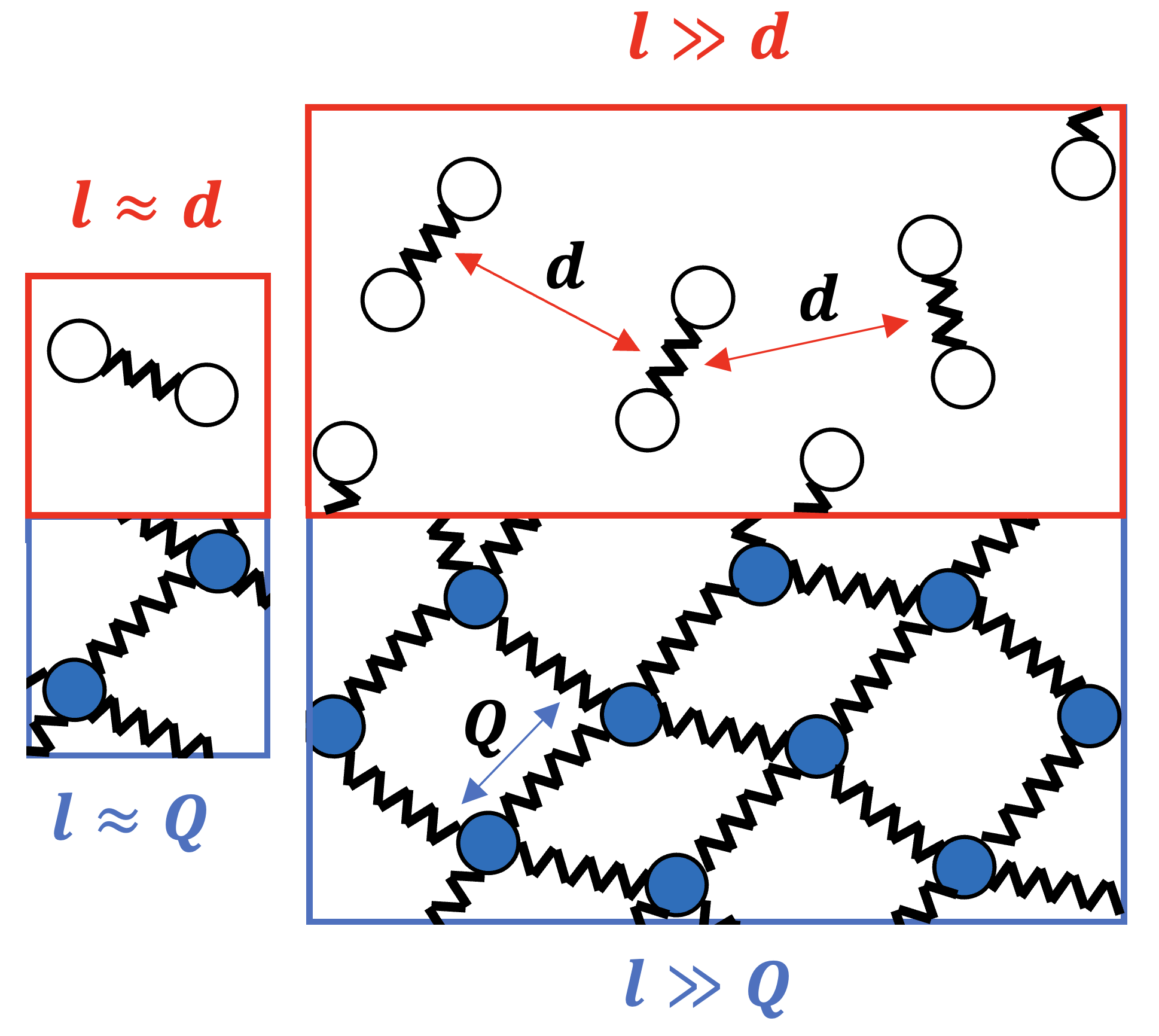}
 \caption{The difference in $\langle\delta\pmb{c}^2\rangle$ is maximal on the scale of \emph{one} structural variable and becomes indistinguishable with increasing number of dumbbells/strands $N_Q$ scaling as in eq. (\ref{eq: M_2(Tau)}). The cartoon depicts how $N_Q$ i.e. the volume $V=N_Q/n$ relates to different characteristics lengths for the two systems.} 
 \label{fig: volumeseffect}%
\end{figure}

In fluctuating hydrodynamics the relation for the fluctuations of a field variable $\delta f(\pmb{x},\pmb{t})$ and the volume $\Delta V$ in which the latter is measured is
\begin{equation}\label{eq: fluctuacting hydrodynamic}
    \delta f_V(t)=\frac{1}{\Delta V} \int_{\Delta V}\delta f(\pmb{x},t)\d^3x. 
\end{equation}
The noise, normalized by $\Delta V$, is Gaussian, independent of  the volume of observation.\cite{Sengers} This same relation is recovered here for the  HDM,\cite{Huetter2020,Beyond} a cartoon representing the physical picture is given in fig. \ref{fig: volumeseffect}. 
In contrast, sec. \ref{sec: Potentials} points out that for ensembles confined to $N=1$ the TNM's fluctuations are Poissonian. On small length scales they cannot be described by Gaussian noise.\\ 

How do these findings relate to stress measurements? The origin of stress fluctuations in these complex materials is the evolution of $\pmb{c}$ (see eq. (\ref{eq: Kramer})). Thus, in homogeneous flow fluctuations $\delta{\pmb{c}}=\langle\pmb{c}\rangle-\pmb{c}$ directly reflect the different statics of the HDM and the TNM.
When the control volume increases (right-hand side of fig. \ref{fig: volumeseffect})), $\pmb{c}$ results from the sum of independent $\pmb{Q}$'s contributions. Consequently, according to the central limit theorem, larger volumes lead to indistinguishable Gaussian noise. To quantify the convergence to the same noise, we define $N_Q$ as the amount of $\pmb{Q}_k$ strands/dumbbells in $\Delta V$, where $N_Q$ linearly relates to $\Delta V$ through the number density $n$. A local measurement for a larger $\Delta V$, as in the right-hand picture of fig. \ref{fig: volumeseffect}, consequently results in $\pmb{c}=\langle\pmb{Q}_k^2\rangle_{N_Q}$. The averages of $\pmb{c}$ over $N_c$ measurements yield
\begin{equation}\label{eq: M_2(Tau)}
   \begin{split}
   \langle \pmb{c}^2 \rangle_{N_c} &= \langle \langle \pmb{Q}\pmb{Q}\rangle_{N_Q}^2 \rangle_{N_c} =\frac{1}{N_c N_Q^2} \sum_{j=1}^{N_c}\big( \sum_{k=1}^{N_Q} \pmb{Q}_k\pmb{Q}_k \big)^2\\
   &= \frac{1}{N_Q}\langle\pmb{Q}\pmb{Q}\pmb{Q}\pmb{Q}\rangle+\frac{N_Q-1}{N_Q}\langle\pmb{Q}_m^2\pmb{Q}_l^2\rangle\\
   &= \frac{1}{N_Q}\langle\pmb{Q}\pmb{Q}\pmb{Q}\pmb{Q}\rangle+\frac{N_Q-1}{N_Q}\pmb{c}^2,
\end{split}
\end{equation}
where we applied binomial coefficients and the independence of $\pmb{Q}_m$ and $\pmb{Q}_l$. Thence, the difference $\langle\delta{\pmb{c}}_{\text{TNM}}^2\rangle-\langle\delta{\pmb{c}}_{\text{HDM}}^2\rangle$ scales with $1/N_Q$.

In agreement with the findings of sec. \ref{sec: Potentials}, when one trajectory is considered (left-hand side of fig. \ref{fig: volumeseffect}) the difference is maximal and corresponds to the TNM's fourth cumulant $C_{ijkl}^{\text{TNM}}=\mu_{ijkl}^{\text{TNM}}-\mu_{ijkl}^{\text{HDM}}$. While in the HDM the critical length scale associated to one trajectory is given by the average distance $d$, in the case of the TNM we have $d=|\pmb{Q}|$.\\

The measurement of the different $\delta\pmb{c}$ in the two models, requires a certain amount of data points. Numerical simulations of the two models quantify exactly how many. The error in the 4\textsuperscript{th} moment $\mu_{ijkl}$ is
\begin{equation} \label{eq: error}
    \epsilon_{ijklijkl}=\sqrt{\frac{\mu_{ijklijkl}-\mu_{ijkl}^2}{N_c-1}},
\end{equation}
so that for a detectable difference we need 
\begin{equation}\label{eq: Condition}
    {\epsilon}^{\text{HDM}}_{ijklijkl}+{\epsilon}^{\text{TNM}}_{ijklijkl}<<C_{ijkl}^{\text{TNM}}
\end{equation}
to hold. Interestingly, an application of these formulas to the analytical solution for the step-strain example (\ref{fig: PDFs}) shows $N_c$ to be independent from the Maxwell parameters $G,\lambda$, but strongly dependent on $\gamma$ with a leading order of $\mathcal{O}(\gamma^{-4})$. Differences in $\delta\pmb{c}$ are most easily to observe far from equilibrium conditions as suggested by the potential in sec. \ref{sec: Potentials} and fig. \ref{fig: PDFs}.

\section{Discussion} \label{sec: Discussion}
\subsection{Relevance of stress fluctuations}
Fluctuations are pivotal for \emph{modeling coupled phenomena} that a purely continuum mechanical description cannot intuitively capture. Examples include reaction-diffusion systems, where fluctuations explain phenomena such as pattern formations,\cite{wang2007pattern} fluctuation-induced instabilities,\cite{kessler1998fluctuation} front propagation dynamics,\cite{khain2011fluctuations} and the emergence of new steady states.\cite{togashi2004molecular} Fluctuating hydrodynamics (\ref{eq: Fluct_hydrodynamics}) with Gaussian noise proved able to capture many of these effects in Newtonian fluids.\cite{Donev} For networks at specific length scales, however, we need the noise description introduced in the present work. The noise enhancement of a UCM of \citet{Huetter2020} is only a \emph{particular} case reflecting the dynamics of dumbbells.\ 

\indent
Likewise, the results are relevant for \emph{interpreting experiments} involving diffusion coupled with non-zero stresses. An illustrative example of this application is observed in the TNM behavior of stretched fire ant networks.\cite{Vernerey2023} In this scenario, the structural variable $\pmb{Q}$ extends as long as the ants' legs. Hence, diffusive phenomena in these stretched networks shall reflect the Poissonian fluctuations dynamics. Likewise, microrheology measurements of a bead embedded in a network under deformation \cite{GOREN2024, Hasnain2006} necessitate a deeper understanding of how Brownian forces change relative to thermodynamic equilibrium. We demonstrated that even in a simple network model like the TNM, the presence of stresses alters the distribution of stochastic forces (see sec. \ref{subsec: measure} and fig. \ref{fig: Structural to bead}).\

\indent
This study also underscores the importance of \emph{stress fluctuations ($\delta \pmb{c},\, \delta \pmb{\tau}$) as indicators of the underlying structure}. Theoretical studies and simulations looking into microscopic mechanisms governing the rheological behavior of colloidal networks by analyzing spontaneous microscopic dynamics, confirming a relation between microstructural dynamics and microscopic structure.\cite{Del_Gado_Nat_2023} Similarly, stress correlations $\langle\delta\pmb{\tau}^2\rangle$, albeit in spatial dimensions rather than temporal, were used to identify force chains in colloidal networks, underscoring the significance of this metric in characterizing microstructures.\cite{Del_Gado_JCP} Recent experiments and simulations have also observed non-Gaussian fluctuations near the glass transition of glass-forming liquids.\cite{Greco} These studies highlight the importance of changes of the temporal displacement distributions in networks with increasing distance from equilibrium. We therefore suggest the possible use of deviations from Gaussian Statistics in stress fluctuations as a means to discern local loads in viscoelastic networks. Such an approach may enable the detection of local stress concentrations.
The present paper identifies the general FDT formulation (Eq. (\ref{eq: General FDT})) as a suitable framework to consistently coarse-grain such noise, offering a detailed examination of fluctuations in a mesoscopic network representation (TNM) for microscopic reference.
\subsection{Experimental determination of the difference in $\delta\pmb{c}$} \label{subsec: measure}
\begin{figure}
  \centering
  {\includegraphics[width=0.9\linewidth]{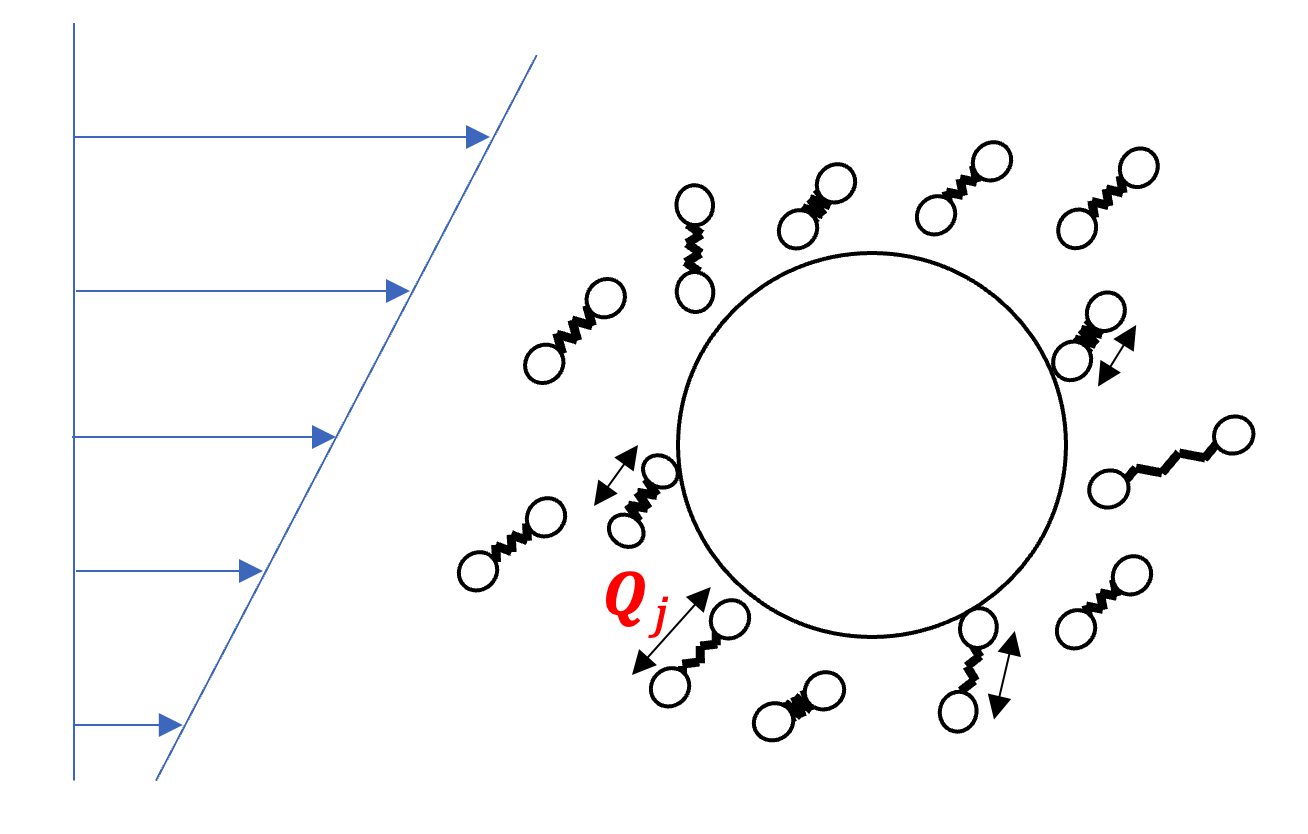}}%
  \caption{The fluctuations of a hard sphere immersed in a fluid during flow reflect the fluctuations of the fluid's structure through its structural variable $\pmb{Q}$.}
  \label{fig: Structural to bead}
\end{figure}

To detect the differences in $\delta\pmb{c}$ and, thus, in the stress $\pmb{\tau}$ (eq. (\ref{eq: Kramer})), several experimental methods may be : \emph{microrheology, Raman scattering, polarized fluorescence, polarimetry}.\\
In \emph{microrheology}  the  mean square displacement of small colloids ($0.1$ to $10$ \textmu m) suspended in a fluid is measured via microscopy or light scattering and related to the linear viscoelastic fluid properties.\cite{Mason1995_1, Savin,Zanten,HE2010141,Furst} However, the techniques relies on an equilibrium formulation. Fig. \ref{fig: Rheoconf} shows that experimental acquisition of the trajectories under flow is now possible. Fig. \ref{fig: Structural to bead} illustrates the origin of the bead's fluctuations - namely the colloid's interaction with the fluids structure, as captured by the structural variable $\pmb{Q}$. The bead's trajectories hence should reflect the distinct nature of fluctuations in the TNM and the HDM in flow.\\
\emph{Raman scattering} defines the frequency shift in a small portion of scattered light due to vibrational and rotational motions within a sample.\cite{Fuller_book} The Raman tensor $\alpha_{ij}$ establishes a connection between the electric vector of incident light and Raman scattered light. Since $\alpha_{ij}$ is a function of segment vectors $\pmb{Q}$, it enables both 2\textsuperscript{nd} and 4\textsuperscript{th} moments to be determined (see, e.g., eq. (6)-(8) in \citet{ARCHER92}). The method has been successfully employed for the quantitative description of $\langle\pmb{Q}\pmb{Q}\rangle$ and $\langle\pmb{Q}\pmb{Q}\pmb{Q}\pmb{Q}\rangle$ for LC systems,\cite{Archer_96_LC} for elongated or sheared polymer melts,\cite{Bower72, Purvis97, Archer_94, Archer_96_pol} and for \emph{in situ} observation of polymerization in a rheometer.\cite{Brun2013} In \citet{Archer_94}, anisotropies were measured for an entangled polyisobutylene melt during step uniaxial extension and compared with predictions from different models, including the TNM. Consequently, Raman scattering has already proven to be a valuable tool to confirm the difference in $\delta\pmb{c}$, but possibly time resolution can be improved.\\
Similarly, \emph{polarized fluorescence} provides the 2\textsuperscript{nd} and 4\textsuperscript{th} moments. It relies on the different wavelengths of emitted and incident light\cite{Fuller_book, Larson} and  has been successfully applied to polymer dynamics.\cite{Monnerie_87,Monnerie_89,Monnerie_94}\\
Linear birefringence using \emph{polarimetry} uses the dependency of refraction from segmental orientation. The polarization by an electric field (such as a light beam) results in a refractive index tensor $\pmb{n}$. The latter is then linearly related to $\pmb{c}$ through the stress-optical law.\cite{Larson,Janeschitz_83} Full tensor optical rheometry was able to determine $\pmb{c}$ in polymeric flows\cite{Kalogrianitis96} and may also be extended to detect  differences in $\delta\pmb{c}$ (\ref{eq: M_2(Tau)}) through point-wise measurements (i.e. using cross correlation techniques for beams sent at slightly different angles).

\section{Summary} \label{sec: Conclusions}
Fluctuations play a crucial role in understanding various processes, spanning from phenomena such as diffusion to slip, and they must be accounted for in order to accurately model coupled phenomena. Furthermore, even though often perceived merely as noise, fluctuations directly probe the material structure and in a unique manner, thereby uncovering the coupling of local dynamics and structure. However, our understanding of fluctuations is largely based on fluctuating hydrodynamics, which describes Newtonian fluids assuming Gaussian noise.\\

The central message of this paper is that the Gaussian approximation does not always hold in complex fluids.  Gaussian noise, as assumed in fluctuating hydrodynamics or in the widely used Fluctuation-Dissipation Theorem formulation of Green-Kubo (eq. (\ref{eq: Green-Cubo})), cannot describe the noise at the length scale of $\pmb{Q}$ in a temporary network model. These findings serve as a caveat, that fluctuations in complex flows may differ fundamentally from those in Newtonian fluids. We introduce the general formulation of the Fluctuation-Dissipation Theorem (\ref{eq: General FDT}) as an effective solution to this problem. This formulation can  account for the discrete jump dynamics occurring in networks, making it a potent tool for developing a fluctuating rheology akin to fluctuating hydrodynamics. \\

Second, Whereas the stress evolutions for dumbbell model and the temporary network model are identical, the stress fluctuations do pick up the differences in microstructure flow coupling. As measuring techniques advance, there is a shift towards less coarse-grained descriptions in rheology. We demonstrate how, within the context of local field theory, the most fundamental quasi-linear viscoelastic model (UCM) exhibits varying fluctuations depending on the underlying microscopic dynamics.\\

Finally, we provide quantification of these differences in stress fluctuations. We suggest using the 4th moment to capture differences in $\delta\pmb{c}$, introduce an analytical solution to a step strain example, and discuss the convergence behavior of $\delta\pmb{c}$ with increasing control volumes. Finally, we emphasize the relevance of these findings for current research and list techniques to detect these differences.

\begin{acknowledgments}
Thanks to  Alberto Montefusco for the discussions on his seminal work.\cite{Alberto_thesis} Thanks to Vincenzo Ianniello, Florence Müller and Martin Kröger (ETH Zürich) for precious feedback about the work. At last, thanks to Pierre Lehéricey (ETH Zürich) for helping with the experiments to produce fig. \ref{fig: Rheoconf}.
\end{acknowledgments}

\newpage

\section*{References}
\bibliography{refs}

\end{document}